\documentclass{achemso}

\usepackage{amssymb}
\usepackage{amsthm}
\usepackage{amsmath}
\usepackage{braket}
\usepackage{graphicx}
\usepackage{hyperref}

\setkeys{acs}{maxauthors=0}

\SectionNumbersOn

\author{Victor Wen-zhe Yu}
\affiliation{Materials Science Division, Argonne National Laboratory, Lemont, Illinois 60439, United States}

\author{Yu Jin}
\affiliation{Pritzker School of Molecular Engineering, The University of Chicago, Chicago, Illinois 60637, United States}

\author{Giulia Galli}
\email{gagalli@uchicago.edu}
\affiliation{Pritzker School of Molecular Engineering, The University of Chicago, Chicago, Illinois 60637, United States}
\alsoaffiliation{Department of Chemistry, The University of Chicago, Chicago, Illinois 60637, United States}
\alsoaffiliation{Materials Science Division, Argonne National Laboratory, Lemont, Illinois 60439, United States}

\author{Marco Govoni}
\email{mgovoni@unimore.it}
\affiliation{Department of Physics, Computer Science, and Mathematics, University of Modena and Reggio Emilia, Modena, 41125, Italy}
\alsoaffiliation{Materials Science Division, Argonne National Laboratory, Lemont, Illinois 60439, United States}
\alsoaffiliation{Pritzker School of Molecular Engineering, The University of Chicago, Chicago, Illinois 60637, United States}

\title{GPU-Accelerated Solution of the Bethe-Salpeter Equation for Large and Heterogeneous Systems}


\begin{document}

\begin{abstract}
We present a massively parallel, GPU-accelerated implementation of the Bethe-Salpeter equation (BSE) for the calculation of the vertical excitation energies (VEEs) and optical absorption spectra of condensed and molecular systems, starting from single-particle eigenvalues and eigenvectors obtained with density functional theory. The algorithms adopted here circumvent the slowly converging sums over empty and occupied states and the inversion of large dielectric matrices, through a density matrix perturbation theory approach and a low-rank decomposition of the screened Coulomb interaction, respectively. Further computational savings are achieved by exploiting the nearsightedness of the density matrix of semiconductors and insulators to reduce the number of screened Coulomb integrals. We scale our calculations to thousands of GPUs with a hierarchical loop and data distribution strategy. The efficacy of our method is demonstrated by computing the VEEs of several spin defects in wide-band-gap materials, showing that supercells with up to 1000 atoms are necessary to obtain converged results. We discuss the validity of the common approximation that solves the BSE with truncated sums over empty and occupied states. We then apply our GW-BSE implementation to a diamond lattice with 1727 atoms to study the symmetry breaking of triplet states caused by the interaction of a point defect with an extended line defect.
\end{abstract}

\section{Introduction}
\label{sec:introduction}

Predicting excited-state properties of molecules and materials from first principles is critical for various applications in photovoltaics, photocatalysis, optoelectronics, microelectronics, quantum technology, and other areas of physics, chemistry, and materials science. Solving the Bethe-Salpeter equation (BSE)~\cite{bse_salpeter_1951,bse_albrecht_1998,bse_onida_2002} within the framework of many-body perturbation theory (MBPT), starting from the output of a density functional theory (DFT) calculation, has proven to be successful in simulating excited states of isolated and periodic systems. However, conventional BSE solvers construct a two-body exciton Hamiltonian in an electron-hole (e-h) basis set, and the computational cost of its diagonalization scales as $\mathcal{O}(N^6)$, where $N$ is the number of electrons in the system. This unfavorable scaling makes it challenging to solve the BSE for heterogeneous materials represented by large supercells, due to the difficulty in calculating multiple screened Coulomb integrals between occupied and empty single-particle states. The implementations of the BSE based on exciton Hamiltonians require truncating sums over empty and occupied states, leading to numerical approximations that are often difficult to control, especially for large systems.

Several techniques have been developed to address the computational challenges of solving the BSE. Rocca et al.~\cite{bse_rocca_2010,bse_rocca_2012,bse_ping_2012} solved the BSE using the linearized Liouville equation and density matrix perturbation theory to circumvent the explicit computation of virtual states, a technique that was earlier applied to time-dependent density functional theory (TDDFT)~\cite{lanczos_walker_2006,lanczos_rocca_2008,turbotddft_malcioglu_2011}. Nguyen et al.~\cite{bse_nguyen_2019} used instead a finite-field approach that directly evaluates the screened Coulomb integrals by solving the Kohn-Sham (KS) equations~\cite{dft_hohenberg_1964,dft_kohn_1965} in a finite electric field, thus allowing for calculations beyond the random phase approximation (RPA). Elliott et al.~\cite{bse_elliott_2019} accelerated two computationally demanding steps in solving the BSE by bypassing the computation of quasi-particle (QP) energies and the direct calculation of the screened Coulomb interaction. Such acceleration was achieved through the use of Koopmans-compliant functionals and a direct minimization scheme applied within a maximally localized Wannier function basis. Dong et al.~\cite{bse_dong_2021} proposed a data-driven model for the dielectric screening that was used to solve the BSE. By learning a mapping from the unscreened to the screened Coulomb integrals, Dong et al. achieved a speedup of over one order of magnitude when computing ultraviolet-visible (UV/vis) absorption spectra at finite temperature, obtained by averaging over multiple snapshots extracted from first-principles molecular dynamics trajectories. All these works adopted techniques that reduce the computational cost of solving the BSE for large systems by avoiding the explicit computation of virtual states. Further computational savings may be obtained by exploiting the nearsightedness of the density matrix of semiconductors and insulators. Instead of using the occupied eigenstates of the KS equations, a set of localized orbitals is obtained e.g., by Wannier localization~\cite{bse_marsili_2017,bse_merkel_2024} or the recursive subspace bisection~\cite{bse_nguyen_2019} method, and the evaluation of screened Coulomb integrals is restricted to those localized wave functions that overlap with each other. Instead of using deterministic methods, Bradbury et al.~\cite{bse_bradbury_2022} used a stochastic time-dependent Hartree propagation technique to calculate the screened Coulomb interaction. The computational complexity was reduced by propagating only a few stochastic orbitals instead of all the occupied ones.

Overall, methodological and algorithmic developments of the past decade have enabled the solution of the BSE for systems with up to a thousand atoms~\cite{bse_nguyen_2019,bse_dong_2021,bse_marsili_2017,adf_forster_2022}. However, tackling larger and more complex systems still poses significant challenges. Despite the adoption of graphics processing units (GPUs) in high-performance computing (HPC), which provides a promising pathway to large-scale BSE calculations, and despite the use of GPU acceleration in a variety of electronic structure software packages~\cite{roadmap_gavini_2023,bigdft_ratcliff_2018,yambo_sangalli_2019,abinit_romero_2020,berkeleygw_delben_2020,cp2k_kuhne_2020,fhiaims_huhn_2020,nwchem_apra_2020,octopus_tancognedejean_2020,onetep_prentice_2020,qe_giannozzi_2020,siesta_garcia_2020,terachem_seritan_2020,turbomole_balasubramani_2020,votcaxtp_tirimbo_2020,inq_andrade_2021,west_yu_2022}, GPU-accelerated BSE calculations have been scarce in the literature. Zhang et al.~\cite{bse_zhang_2021} reported a GPU-accelerated diagonalization of the BSE Hamiltonian constructed on central processing units (CPUs). Further, Franzke et al.~\cite{bse_franzke_2022} conducted GPU-accelerated BSE calculations to compute nuclear magnetic resonance properties of molecules. However, to the best of our knowledge, a massively parallel, scalable, and GPU-accelerated implementation of a BSE solver for systems consisting of more than a thousand atoms has not yet been reported.

In this paper, we present GW-BSE calculations of unprecedented size, powered by a massively parallel and GPU-accelerated BSE solver implemented in the open-source WEST (Without Empty STates) code~\cite{west_yu_2022,west_govoni_2015,west_website}. WEST features a plane-wave pseudopotential implementation of MBPT, including full-frequency G$_0$W$_0$~\cite{west_yu_2022,west_govoni_2015,soc_scherpelz_2016,gw100_govoni_2018,gw_ma_2019} and electron-phonon self-energy~\cite{phonon_mcavoy_2018,phonon_yang_2021,phonon_yang_2022} calculations; the solution of the BSE~\cite{bse_nguyen_2019,bse_dong_2021}; the calculation of excited-state energies and forces with (spin-flip) TDDFT~\cite{tddft_jin_2022,tddft_jin_2023}; and the calculations of vertical excitation energies (VEEs) between multireference states using the quantum defect embedding theory (QDET)~\cite{qdet_ma_2020a,qdet_ma_2020b,qdet_ma_2021,qdet_sheng_2022}. The algorithms adopted in WEST are specifically designed to avoid or alleviate computational and memory bottlenecks in simulations of large systems. For instance, the slowly converging sum over virtual states, commonly encountered in most MBPT codes, is completely sidestepped. Moreover, WEST features a multilevel parallelization scheme that fully leverages the embarrassingly parallel parts of the algorithms adopted in the code. WEST has been demonstrated to scale the calculation of full-frequency G$_0$W$_0$ self-energies to 524288 CPU cores~\cite{west_govoni_2015} and to 25920 GPUs~\cite{west_yu_2022}. Here we report the GPU acceleration of the BSE solver (hereafter referred to as WEST-BSE) and its excellent performance and scalability on GPU-equipped HPC systems. As examples of representative systems, we computed the VEEs of spin defects in large supercells of diamond and silicon carbide (SiC), obtaining results that are converged with respect to the supercell size. In addition, we studied the nitrogen-vacancy center at a dislocation core in diamond, showcasing the capability to investigate, at the MBPT level of theory, the interaction between a point defect and an extended defect.

The rest of the paper is organized as follows. Section~\ref{sec:theory} reviews the BSE formulation implemented in the WEST code. Section~\ref{sec:gpu} describes the implementation of the WEST-BSE code on GPUs. In section~\ref{sec:performance}, we present performance benchmarks of WEST-BSE on leadership HPC systems, demonstrating its scalability to 4096 GPUs. Section~\ref{sec:applications} reports large-scale GW-BSE calculations for defective systems consisting of thousands of electrons. Our concluding remarks are presented in section~\ref{sec:conclusions}.

\section{Theory}
\label{sec:theory}

The solution of the BSE is implemented in a number of electronic structure codes: ABINIT~\cite{abinit_romero_2020}, BerkeleyGW~\cite{berkeleygw_deslippe_2012}, GPAW~\cite{gpaw_huser_2013}, GWL~\cite{bse_marsili_2017}, VASP~\cite{vasp_kresse_1996}, WEST~\cite{west_govoni_2015}, and Yambo~\cite{yambo_sangalli_2019} using plane-wave basis sets, CP2K~\cite{cp2k_kuhne_2020} using mixed plane-wave and localized basis sets, Elk~\cite{elk} and Exciting~\cite{exciting_gulans_2014} using linearized augmented-plane-waves with local orbitals, Fiesta~\cite{fiesta_blase_2011}, MOLGW~\cite{molgw_bruneval_2016}, TURBOMOLE~\cite{turbomole_balasubramani_2020}, and VOTCA-XTP~\cite{votcaxtp_tirimbo_2020} using Gaussian basis sets, ADF~\cite{adf_forster_2022} using Slater type orbitals, FHI-aims~\cite{fhiaims_liu_2020} using numerical atomic orbitals, and NanoGW~\cite{nanogw_tiago_2006} using real-space grids. The conventional BSE formulation builds a two-body exciton Hamiltonian in an e-h basis set, formed by the product of occupied and empty states. Therefore, the size of the basis set scales as $\mathcal{O}(N^2)$, and the computational complexity of directly diagonalizing the BSE Hamiltonian scales as $\mathcal{O}(N^6)$. In order to reduce the computational cost, the BSE is often solved within a predefined energy window that includes a chosen number of the highest occupied states and lowest empty states. Depending on the system, obtaining converged excitation energies with respect to the number of states or computing the optical absorption spectrum across a broad energy range may be challenging.

We now turn to describing the BSE formulation employed in the WEST code, which is built upon the linearized Liouville equation and density matrix perturbation theory~\cite{lanczos_rocca_2008,bse_rocca_2010,bse_rocca_2012,bse_ping_2012}. As we show below, in WEST-BSE the definition of an energy window is not necessary. In WEST, BSE calculations are implemented by sampling the Brillouin zone at the $\Gamma$ point of the supercell; the latter is an appropriate choice for large supercells and for systems where the periodicity is broken, e.g., solids with defects or interfaces between different condensed systems. The code also supports spin-polarized calculations, including spin-conserving and spin-flip~\cite{tddft_jin_2023} options. For simplicity, the spin and $k$-point indices are omitted hereafter.

Within the Tamm-Dancoff approximation (TDA)~\cite{tda_hirata_1999}, the VEE $\omega_s$ from the ground to the $s$-th excited state can be computed by solving the following eigenvalue problem:
\begin{equation}
\label{eq:evp}
(D + K^{1e} - K^{1d}) A_s = \omega_s A_s \,,
\end{equation}
where $(D + K^{1e} - K^{1d})$ is the Liouville operator, and $A_s = \{ \ket{a_{s,v}}: v = 1, \dots, N_{\mathrm{occ}} \}$ denotes a set of auxiliary orbitals that enter the definition of the linear change of the density matrix with respect to the ground state, due to the $s$-th neutral excitation:
\begin{equation}
\Delta \rho_s = \sum_{v = 1}^{N_{\mathrm{occ}}} \ket{a_{s,v}} \bra{\psi_v} \,,
\end{equation}
where $N_{\mathrm{occ}}$ is the number of occupied states, and $\ket{\psi_v}$ are the KS wave functions of the ground state. The $D$, $K^{1e}$, and $K^{1d}$ terms in equation~\ref{eq:evp} are defined as follows:
\begin{equation}
\label{eq:d}
D A_s = \left\{ \hat{P}_c (\hat{H}^o_{\mathrm{QP}} - \varepsilon_v^{\mathrm{QP}}) \ket{a_{s,v}} : v = 1, \dots, N_{\mathrm{occ}} \right\} \,,
\end{equation}
\begin{equation}
\label{eq:k1e}
K^{1e} A_s = \left\{ 2 \int \mathrm{d} \mathbf{r}' \hat{P}_c (\mathbf{r}, \mathbf{r}') \psi_v (\mathbf{r}') \sum_{v' = 1}^{N_{\mathrm{occ}}} \int \mathrm{d} \mathbf{r}'' v_c (\mathbf{r}', \mathbf{r}'') \psi_{v'}^* (\mathbf{r}'') a_{s,v'} (\mathbf{r}'') : v = 1, \dots, N_{\mathrm{occ}} \right\} \,,
\end{equation}
\begin{equation}
\label{eq:k1d}
K^{1d} A_s = \left\{ \int \mathrm{d} \mathbf{r}' \hat{P}_c (\mathbf{r}, \mathbf{r}') \sum_{v' = 1}^{N_{\mathrm{occ}}} \tau_{vv'} (\mathbf{r'}) a_{s,v'} (\mathbf{r'}) : v = 1, \dots, N_{\mathrm{occ}} \right\} \,,
\end{equation}
where $\varepsilon_v^{\mathrm{QP}}$ is the QP energy corresponding to the $v$-th wave function $\psi_v$, $\hat{P}_c = 1 - \sum_{v = 1}^{N_{\mathrm{occ}}} \ket{\psi_v} \bra{\psi_v}$ is the projector onto the unoccupied subspace, $\hat{H}^o_{\mathrm{QP}}$ is the QP Hamiltonian of the ground state, $v_c (\mathbf{r}, \mathbf{r'}) = (|\mathbf{r} - \mathbf{r'}|)^{-1}$ is the bare Coulomb potential, and $\tau_{vv'} (\mathbf{r})$ is the screened Coulomb integral between states $v$ and $v'$,
\begin{equation}
\label{eq:tau}
\tau_{vv'}(\mathbf{r}) = \int \mathrm{d}\mathbf{r}' W (\mathbf{r}, \mathbf{r}') \psi_{v} (\mathbf{r}') \psi_{v'}^* (\mathbf{r}') \,,
\end{equation}
where $W$ denotes the statically screened Coulomb interaction. The QP Hamiltonian $\hat{H}^o_{\mathrm{QP}}$ in equation~\ref{eq:d} can be written as
\begin{equation}
\label{eq:ham}
\hat{H}^o_{\mathrm{QP}} = \hat{H}^o_{\mathrm{KS}} + \sum_{n = 1}^{N_{\mathrm{QP}}} \ket{\psi_n} (\varepsilon_n^{\mathrm{QP}} - \varepsilon_n - \Delta \varepsilon) \bra{\psi_n} + \Delta \varepsilon \hat{I} \,,
\end{equation}
where $\hat{H}^o_{\mathrm{KS}}$ is the KS Hamiltonian. Here, we explicitly compute the QP energy corrections for the lowest $N_{\mathrm{QP}} > N_{\mathrm{occ}}$ states (i.e., for $n \leq N_{\mathrm{QP}}$), while for virtual states with higher energy (i.e., for $n > N_{\mathrm{QP}}$), we approximate the difference between the QP energies $\varepsilon_n^{\mathrm{QP}}$ and the KS energies $\varepsilon_n$ as
\begin{equation}
\Delta \varepsilon = \frac{1}{N_{\mathrm{QP}} - N_{\mathrm{mid}} + 1} \sum_{n = N_{\mathrm{mid}}}^{N_{\mathrm{QP}}} \left(\varepsilon_n^{\mathrm{QP}} - \varepsilon_n \right) \,,
\end{equation}
where $N_{\mathrm{mid}} = (N_{\mathrm{occ}} + N_{\mathrm{QP}}) / 2$. In addition, we approximate the QP wave functions with the KS wave functions.

The evaluation of the screened Coulomb integrals $\tau_{vv'}$ for all pairs of occupied states constitutes a severe computational bottleneck since the computation of $W$ entails the calculation of the dielectric response. In the commonly adopted Adler-Wiser formulation~\cite{dielectric_adler_1962,dielectric_wiser_1963}, the dielectric response is evaluated by summing over virtual states and inverting large dielectric matrices. These operations become prohibitively expensive for large systems. Using the relation $W = v_c + v_c \chi v_c$, $\tau_{vv'}$ can be written as:
\begin{equation}
\label{eq:tau_from_tau_u}
\begin{split}
\tau_{vv'} & = \tau_{vv'}^u + v_c \chi \tau_{vv'}^u \\
& = \tau_{vv'}^u + v_c^{1/2} \bar{\chi} v_c^{-1/2} \tau_{vv'}^u \,,
\end{split}
\end{equation}
where $\tau_{vv'}^u$ is the unscreened Coulomb integral
\begin{equation}
\tau_{vv'}^u = \int \mathrm{d}\mathbf{r}' v_c (\mathbf{r}, \mathbf{r}') \psi_{v} (\mathbf{r}') \psi_{v'}^* (\mathbf{r}') \,,
\end{equation}
and $\chi$ is the density-density response function, and its symmetrized form is $\bar{\chi} = v_c^{1/2} \chi v_c^{1/2}$.

To compute $\bar{\chi}$ at zero frequency without explicitly computing virtual states or inverting large dielectric matrices, we turn to the projective dielectric eigenpotentials (PDEP) technique~\cite{dielectric_wilson_2008,dielectric_wilson_2009,pdep_nguyen_2012,pdep_pham_2013}, which iteratively diagonalizes the symmetrized irreducible density-density response function $\bar{\chi}^0$. Using the relation $\bar{\chi} = \bar{\chi}^0 + \bar{\chi}^0 \bar{\chi}$, which is valid within the RPA, $\bar{\chi}$ can be expressed as
\begin{equation}
\label{eq:pdep}
\bar{\chi} = \Xi + \frac{1}{\Omega} \sum_{a = 1}^{N_{\mathrm{PDEP}}} \ket{\bar{\varphi}_a} \frac{\lambda_a}{1 - \lambda_a} \bra{\bar{\varphi}_a} \,,
\end{equation}
where $\lambda_a$ and $\bar{\varphi}_a$ denote the $a$-th eigenvalue and eigenfunction of $\bar{\chi}^0$, respectively; the $\Xi$ term takes into account the long-range dielectric response, $\Omega$ is the volume of the supercell, and $N_{\mathrm{PDEP}}$ is the number of eigenpotentials that determines the accuracy of the low-rank approximation in equation~\ref{eq:pdep}. Previous benchmarks~\cite{gw100_govoni_2018} suggest that converged results can be obtained with $N_{\mathrm{PDEP}}$ being just a few times the number of electrons. The computational complexity of the PDEP method scales as $N_{\mathrm{occ}}^2 \times N_{\mathrm{PDEP}} \times N_{\mathrm{pw}}$~\cite{west_yu_2022,west_govoni_2015,dielectric_wilson_2008,dielectric_wilson_2009,pdep_nguyen_2012,pdep_pham_2013}, while that of the conventional Adler-Wiser formulation~\cite{dielectric_adler_1962,dielectric_wiser_1963} scales as $N_{\mathrm{occ}} \times N_{\mathrm{empty}} \times N_{\mathrm{pw}}^2$, where $N_{\mathrm{occ}}$ ($N_{\mathrm{empty}}$) is the number of occupied (virtual) states, and $N_{\mathrm{pw}}$ is the number of plane-waves used to represent the wave functions (in practice $N_{\mathrm{pw}} \gg N_{\mathrm{PDEP}}$).

The number of $\tau_{vv'}$ integrals to be evaluated can be reduced by utilizing the nearsightedness of the density matrix of semiconductors and insulators. We introduce a unitary transformation $U$ to localize the occupied KS wave functions:
\begin{equation}
\label{eq:wannier}
\tilde{\psi}_v = \sum_{v' = 1}^{N_{\mathrm{occ}}} U_{vv'} \psi_{v'} \,.
\end{equation}
In this work, the $N_{\mathrm{occ}} \times N_{\mathrm{occ}}$ matrix $U$ is determined using the joint approximate diagonalization of eigen-matrices (JADE)~\cite{wannier_gygi_2003,jade_holobar_2006} algorithm. Specifically, JADE simultaneously diagonalizes, in an approximate manner, six noncommuting real symmetric matrices obtained by representing the operators $M^{(n)}, n = 1, \dots, 6$ on occupied wave functions. For an orthorhombic cell of edges $L_x$, $L_y$, and $L_z$, the operators $M^{(n)}$ are defined as
\begin{equation}
\begin{split}
M^{(1)} = \mathrm{cos} \left( \frac{2 \pi}{L_x} x \right) ,\,
M^{(2)} = \mathrm{sin} \left( \frac{2 \pi}{L_x} x \right) ,\, \\
M^{(3)} = \mathrm{cos} \left( \frac{2 \pi}{L_y} y \right) ,\,
M^{(4)} = \mathrm{sin} \left( \frac{2 \pi}{L_y} y \right) ,\, \\
M^{(5)} = \mathrm{cos} \left( \frac{2 \pi}{L_z} z \right) ,\,
M^{(6)} = \mathrm{sin} \left( \frac{2 \pi}{L_z} z \right) ,\,
\end{split}
\end{equation}
where $x$, $y$, and $z$ are the Cartesian components of the position operators. The general expression of $M^{(n)}$ for cells of arbitrary symmetry was first implemented in Qbox~\cite{qbox_gygi_2008} and specific cases have been discussed earlier in reference~\citenum{wannier_silvestrelli_1999}. The transformation $U$ is determined by minimizing the spread function
\begin{equation}
\sigma^2(U) = \sum_{n = 1}^6 \left[ \mathrm{Tr}\left( U^{\dagger} B^{(n)} U \right) - \sum_{v} \left| \left( U^{\dagger} A^{(n)} U \right)_{vv} \right|^2 \right]
\label{eq:spread}
\end{equation}
with respect to $U$ ($U^{\dagger}$ is the complex conjugate of $U$). The matrices $A^{(n)}$ and $B^{(n)}$ are defined as
\begin{equation}
\begin{split}
A^{(n)}_{vv'} & = \braket{\psi_v | M^{(n)} | \psi_{v'}} \,, \\
B^{(n)}_{vv'} & = \braket{\psi_v | \left( M^{(n)} \right)^2 | \psi_{v'}} \,.
\end{split}
\end{equation}
Because the trace and the Frobenius norm of a matrix are invariant under unitary similarity transformations, the minimization of $\sigma^2(U)$ in equation~\ref{eq:spread} is equivalent to a minimization of the off-diagonal elements of the $A^{(n)}$ matrices, i.e., to a minimization of $\sum_{n,v \neq v'} |(U^\dagger A^{(n)}U)_{vv'}|^2$. In the original algorithm proposed in reference~\citenum{wannier_gygi_2003}, this minimization is carried out by initializing $U$ as the identity matrix. In WEST-BSE, the convergence of the spread minimization is facilitated by initializing $U$ with the eigenvectors of $A^{(1)}$, since the algorithm aims to approximately diagonalize $A^{(n)}, n = 1, \dots, 6$. The JADE algorithm exposes a concurrent loop on $2 \times 2$ Givens rotations that need to be applied to the $U$ and $A^{(n)}$ matrices. To maximize data throughput on a single GPU, we accumulated independent Givens rotations and applied them as a single matrix-matrix multiplication at the end of the concurrent loop. For a silicon supercell with 512 atoms and 2048 occupied bands, the calculation of Wannier orbitals with the JADE algorithm took three minutes on a single NVIDIA A100 GPU.

The overlap between two localized wave functions $\tilde{\psi}_v$ and $\tilde{\psi}_{v'}$ is
\begin{equation}
\label{eq:overlap}
O_{vv'} = \frac{\int \mathrm{d}\mathbf{r} | \tilde{\psi}_v (\mathbf{r}) |^2 | \tilde{\psi}_{v'} (\mathbf{r}) |^2}{\sqrt{\int \mathrm{d}\mathbf{r} | \tilde{\psi}_v (\mathbf{r}) |^4 \int \mathrm{d}\mathbf{r} | \tilde{\psi}_{v'} (\mathbf{r}) |^4}} \,.
\end{equation}
The number of integrals to be computed is reduced by approximating $\tau_{vv'} = 0$ if the corresponding $O_{vv'}$ is smaller than a preset threshold, which is system dependent and can be adjusted to balance the computational cost and the required accuracy of the results~\cite{bse_nguyen_2019,bse_elliott_2019,bse_marsili_2017,tddft_jin_2023}.

Overall, the computational cost of the GW-BSE method presented here scales as $\mathcal{O}(N^4)$, with the evaluation of the screened Coulomb interaction using the PDEP technique (equation~\ref{eq:pdep}) scaling as $\mathcal{O}(N^4)$~\cite{west_yu_2022,west_govoni_2015,dielectric_wilson_2008,dielectric_wilson_2009,pdep_nguyen_2012,pdep_pham_2013} and the application of the Liouville operator (equations~\ref{eq:d}--\ref{eq:k1d}) scaling as $\mathcal{O}(N^3)$~\cite{bse_nguyen_2019,bse_marsili_2017}.

\section{GPU acceleration}
\label{sec:gpu}

\begin{figure}[ht!]
\includegraphics[width=0.3\textwidth]{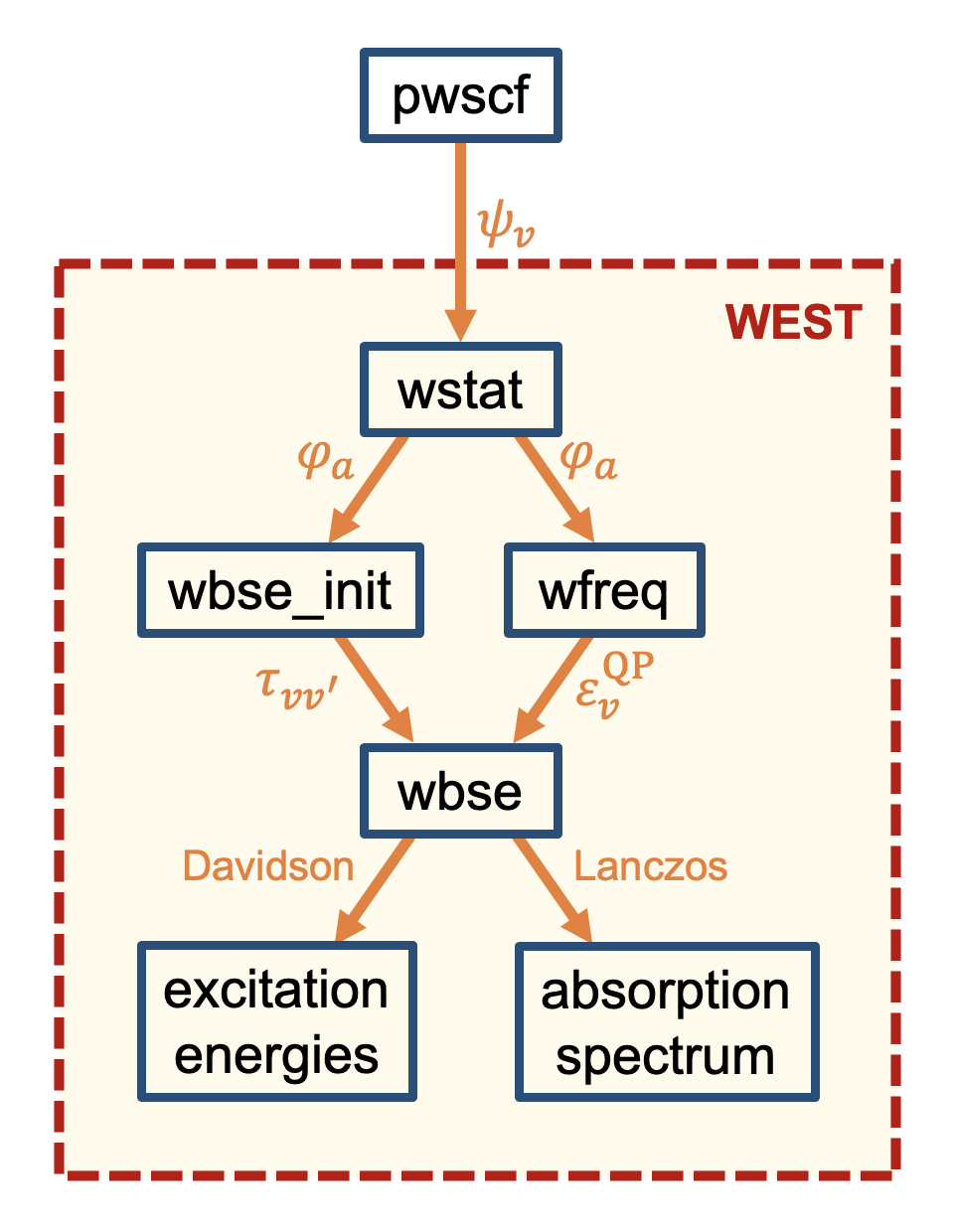}
\caption{Summary of the modules of the WEST-BSE code, used to solve the Bethe-Salpeter equation (BSE). The \texttt{pwscf} code in Quantum ESPRESSO is employed to compute the Kohn-Sham wave functions $\psi_v$, which are read by the \texttt{wstat} module. \texttt{Wstat} diagonalizes the static dielectric matrix to obtain the eigenpotentials $\varphi_a$ (equation~\ref{eq:pdep}). The \texttt{wfreq} and \texttt{wbse\_init} modules use $\varphi_a$ to compute G$_0$W$_0$ quasi-particle (QP) energies $\varepsilon_v^{\mathrm{QP}}$ and screened Coulomb integrals $\tau_{vv'}$ (equation~\ref{eq:tau_from_tau_u}), respectively. Finally, $\varepsilon_v^{\mathrm{QP}}$ and $\tau_{vv'}$ are used by the \texttt{wbse} module to compute excitation energies using equation~\ref{eq:evp} or optical absorption spectra.}
\label{fig:workflow}
\end{figure}

The workflow adopted in the WEST code to solve the BSE is illustrated in figure~\ref{fig:workflow}. All of the modules have been ported to GPUs. The WEST-BSE code adopts a multilevel parallelization strategy that fully leverages the embarrassingly parallel parts of the algorithms presented above~\cite{west_yu_2022,tddft_jin_2023}. First, the processes are partitioned into subgroups, referred to as images, to facilitate the iterative diagonalization of the Liouville operator using the Davidson method~\cite{davidson_davidson_1975}. The latter is carried out by applying the Liouville operator to a set of trial vectors that are distributed to the images. Second, in simulations of spin-polarized systems, the processes within each image are partitioned into two pools, each computing one of the two spin channels. Third, the processes within each pool are further partitioned into band groups, each responsible for computing a subset of single-particle states. Communications between band groups are required only when a summation over occupied states is carried out, e.g., when evaluating equations~\ref{eq:k1e} and \ref{eq:k1d}. Finally, the fast Fourier transforms (FFTs) between the direct and reciprocal spaces and linear algebra operations are carried out within a band group. A schematic visualization of the multilevel parallelization is presented in figure 2, reference \citenum{west_yu_2022}.

\begin{figure*}[ht!]
\includegraphics[width=0.45\textwidth]{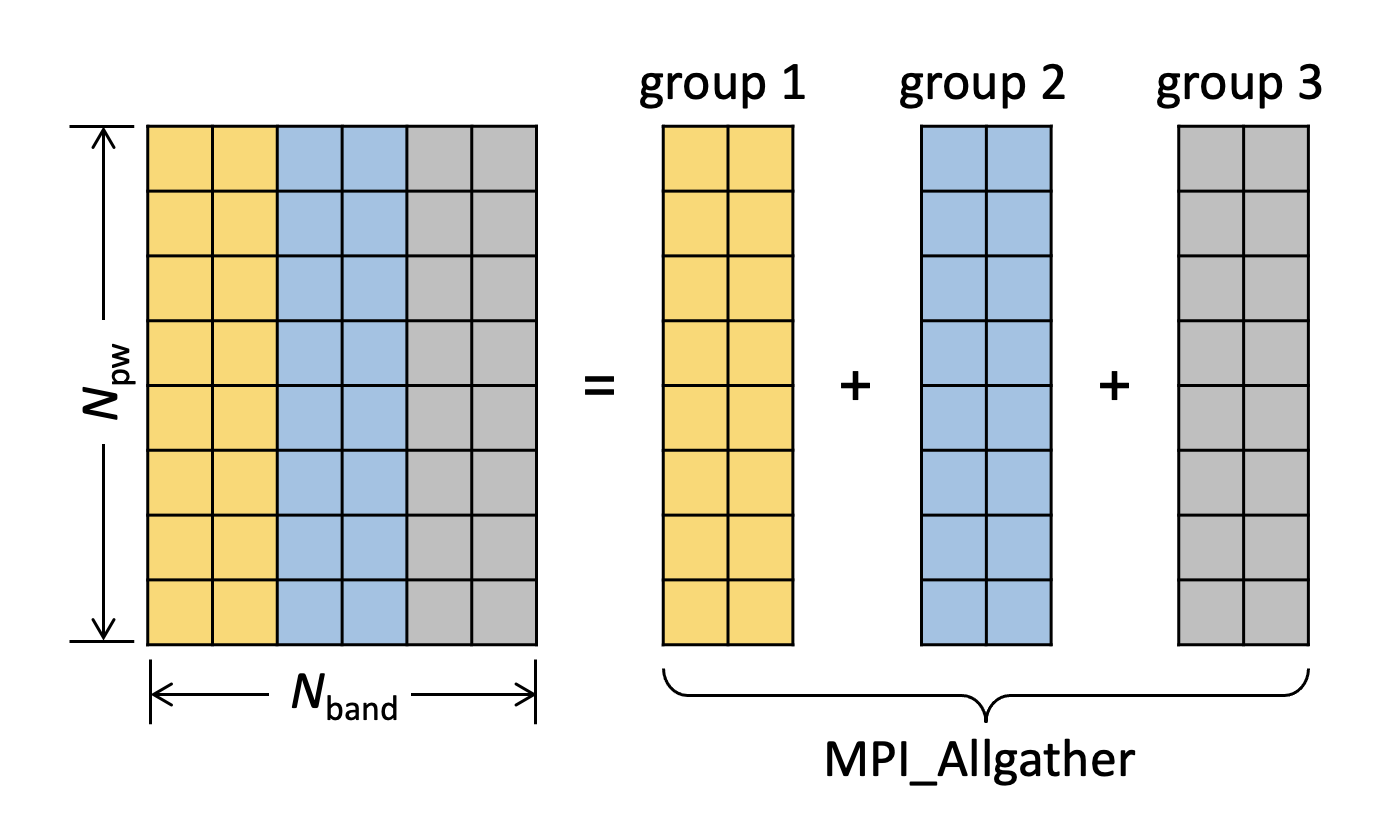}
\caption{Schematic representation of the block data distribution of vectors $A_s$ (see equation~\ref{eq:evp}) across band groups. $N_\mathrm{pw}$ and $N_\mathrm{occ}$ denote the numbers of plane-waves and occupied states, respectively. The summation over band groups is carried out by \texttt{MPI\_Allgather} that gathers data from each band group. Data is color-coded, with data belonging to different band groups shown with different colors.}
\label{fig:distribution}
\end{figure*}

We note that minimizing the communication between GPUs is key to achieving high performance and parallel efficiency. Indeed, as discussed in reference~\citenum{west_yu_2022} for the full-frequency G$_0$W$_0$ calculations of QP energies, an excess use of GPUs within a band group can lead to inefficient parallel FFTs. We hence used the minimum number of GPUs within a band group, that allows for sufficient memory to host the data. In many cases, this choice corresponds to the ideal case of one GPU per band group. Communications across different band groups are required, for instance, to solve equation~\ref{eq:k1d}, where the screened Coulomb integrals $\tau_{vv'}$ are computed in the space of localized Wannier functions, while the vectors $A_s$ are defined in the space of KS wave functions. Thus, a transformation of $A_s$ between the two spaces is necessary and it is obtained by applying the unitary transformation matrix $U$ to the orbitals that define $A_s$, similar to the transformation of wave functions in equation~\ref{eq:wannier}. Note that $A_s$ contains $N_\mathrm{occ}$ orbitals, each of length $N_\mathrm{pw}$. We distributed the $N_\mathrm{occ}$ orbitals across band groups, using a block distribution scheme, as shown in figure~\ref{fig:distribution}. When multiplying the vectors $A_s$ with the transformation matrix $U$, we use the message passing interface (MPI) function \texttt{MPI\_Allgather} to collect all the vectors. This \texttt{MPI\_Allgather} operation is initiated in a non-blocking fashion, allowing for the data communication to proceed while concurrently evaluating the $D$ and $K^{1e}$ terms. We enforce the completion of the communication before evaluating the $K^{1d}$ term. Additionally, when supported by the underlying hardware, we employ GPU-aware MPI to enable direct data transfer between GPUs without accessing the host CPU.

To leverage GPU acceleration, we take advantage of highly optimized third-party GPU libraries, such as cuFFT for FFTs and cuBLAS for basic linear algebra operations on NVIDIA GPUs. Compute loops that cannot use available library functions are offloaded to GPUs through compiler directives, which automatically generate GPU kernels from regions of annotated CPU code. We transitioned from CUDA Fortran, employed in our previous work~\cite{west_yu_2022}, to OpenACC for enhanced functionality and portability, since CUDA Fortran is designed specifically for NVIDIA GPUs. The present version of WEST is compatible with any GPUs for which an OpenACC compiler is available. Work is ongoing to develop and optimize an OpenMP version of the code that is compatible with the Frontier and Aurora exascale supercomputers, equipped with AMD and Intel GPUs, respectively.

\section{Performance and scalability}
\label{sec:performance}

In this section, we report an assessment of the performance and scalability of the WEST-BSE code on the Perlmutter supercomputer at the National Energy Research Scientific Computing Center (NERSC). Each node of the GPU partition of Perlmutter has one AMD EPYC 7763 CPU and four NVIDIA A100 GPUs. We first considered the negatively-charged nitrogen-vacancy center (NV$^-$) in diamond, a prototypical spin defect with numerous quantum technology applications~\cite{nv_doherty_2013,nv_schirhagl_2014,nv_gali_2019,nv_barry_2020}. We computed the VEE from the triplet ground state $^3A_2$ to the triplet excited state $^3E$ of the NV$^-$ in a $5 \times 5 \times 5$ supercell of diamond with 999 atoms and 3998 electrons. We then studied bulk silicon, a typical benchmark system in electronic structure theory. We computed the optical absorption spectrum of an $8 \times 8 \times 8$ supercell of silicon with 1024 atoms and 4096 electrons.

Following the workflow presented in figure~\ref{fig:workflow}, the \texttt{pwscf} code of Quantum ESPRESSO~\cite{qe_giannozzi_2020} was used for all ground-state DFT calculations in this and in the next sections, with the same pseudopotentials, exchange-correlation functional, and numerical parameters as specified below. We employed the SG15 optimized norm-conserving Vanderbilt (ONCV) pseudopotentials~\cite{oncv_hamann_2013,oncv_schlipf_2015} and the PBE exchange-correlation functional~\cite{pbe_perdew_1996}. A kinetic energy cutoff of 60 Ry was used for the plane-wave basis set. The Brillouin zone of the large supercells was sampled with the $\Gamma$-point only. The \texttt{wstat} module of WEST was employed to iteratively diagonalize the static dielectric matrix to obtain the PDEP basis set (equation~\ref{eq:pdep}). The number of the PDEP eigenpotentials was set equal to three and two times the number of electrons for the NV$^-$ and silicon, respectively. Further, the \texttt{wfreq} module of WEST was used to compute the QP energies using the full-frequency G$_0$W$_0$ method~\cite{west_govoni_2015,west_yu_2022}. In addition, the \texttt{wbse\_init} module of WEST was employed to compute the screened Coulomb integrals (equation~\ref{eq:tau}). Wannier localization with the JADE algorithm was enforced to reduce the number of integrals. We set $\tau_{vv'} = 0$ if the overlap of two Wannier functions (equation~\ref{eq:overlap}) falls below 0.001. Finally, the \texttt{wbse} module of WEST was employed to solve the BSE (equation~\ref{eq:evp}). We verified that with the numerical settings specified above, the VEEs of the NV$^-$ and the two peaks at about 3 and 4 eV in the absorption spectrum of Si are converged within 10 meV.

\begin{figure}[ht!]
\includegraphics[width=0.45\textwidth]{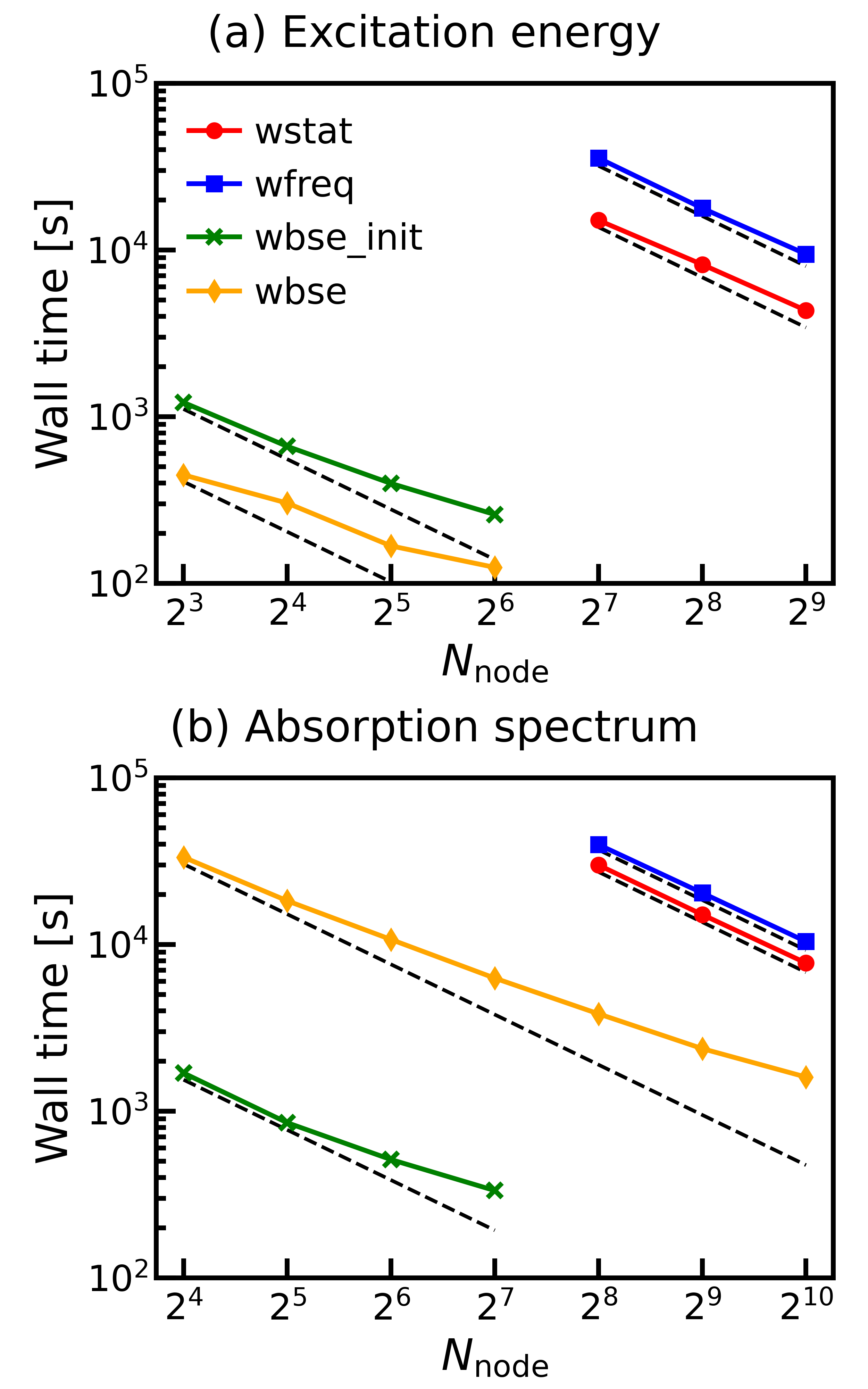}
\caption{Performance of the WEST-BSE code for the calculation of (a) the first vertical excitation energy of the nitrogen-vacancy center in diamond using a $5 \times 5 \times 5$ supercell (3998 electrons) and (b) the optical absorption spectrum of bulk Si using an $8 \times 8 \times 8$ supercell (4096 electrons). Benchmarks were conducted using $N_{\mathrm{node}}$ GPU nodes of the NERSC Perlmutter supercomputer. Timings correspond to the total wall clock time, and include the time spent on I/O operations and communications. The red circles, blue squares, green circles, and orange diamonds correspond to the timings of the \texttt{wstat}, \texttt{wfreq}, \texttt{wbse\_init}, and \texttt{wbse} modules, respectively (see figure~\ref{fig:workflow}). The black dashed lines indicate the slopes of ideal strong scaling.}
\label{fig:scaling}
\end{figure}

Figure~\ref{fig:scaling}(a) reports the performance of the WEST-BSE code to compute the first excitation energy of the NV$^-$ in diamond using a $5 \times 5 \times 5$ supercell. The \texttt{wstat} and \texttt{wfreq} modules scale efficiently to 512 GPU nodes of Perlmutter (2048 GPUs), consistent with our previous benchmarks~\cite{west_yu_2022}. The minor degradation of the parallel efficiency is attributed to the increasingly expensive MPI communications as a function of the number of nodes and to the nonscalable input/output (I/O) operations. The \texttt{wbse\_init} and \texttt{wbse} modules exhibit reasonable scaling to 64 nodes (256 GPUs). The execution of these modules was completed in under 400 seconds using 64 nodes, a time two orders of magnitude smaller than that of \texttt{wstat} and \texttt{wfreq}. Given the relatively low computational costs of \texttt{wbse\_init} and \texttt{wbse}, we did not conduct further benchmarks beyond 64 nodes.

Figure~\ref{fig:scaling}(b) reports the performance of the WEST-BSE code to compute the optical absorption spectrum of bulk Si using an $8 \times 8 \times 8$ supercell. The parallel scaling of the \texttt{wstat} and \texttt{wfreq} modules was tested using up to 1024 GPU nodes of Perlmutter (4096 GPUs). We note that the \texttt{wfreq} step may be skipped by applying a scissor operator to approximate the QP corrections, which is a reasonable approximation for various s-p bonded systems~\cite{gw_hybertsen_1986}. The parallel scaling of the \texttt{wbse\_init} module was tested using up to 128 nodes (512 GPUs). The \texttt{wbse} code exhibits efficient scaling to 1024 nodes (4096 GPUs). The parallel efficiency reported in figure~\ref{fig:scaling}(b) outperforms the one of figure~\ref{fig:scaling}(a) for two key reasons. First, in the case of the diamond $5 \times 5 \times 5$ supercell (volume 5677.4 \AA$^3$), the workload assigned to each GPU becomes relatively low when employing a large number of GPUs. In contrast, in the case of the Si $8 \times 8 \times 8$ supercell (volume 21034.7 \AA$^3$), a substantial amount of work is assigned to each GPU even when employing thousands of GPUs, leading to a higher utilization of each one. Second, the computation of the spectrum requires a significantly larger number of operations than the computation of the lowest excitation energies. The latter are computed using a Davidson algorithm that iteratively diagonalizes the Liouville operator and requires $N_{\mathrm{Davidson}} \times N_s$ times applications of the operator to the vectors $\{A_s: s = 1, \dots N_s\}$, where $N_{\mathrm{Davidson}}$ is the number of iterations required to converge the iterative diagonalization, and $N_s$ is the desired number of VEEs. In figure~\ref{fig:scaling}(a), $N_{\mathrm{Davidson}}$ is 24. The spectrum is instead computed using a Lanczos algorithm that avoids any iterative diagonalization~\cite{lanczos_walker_2006,lanczos_rocca_2008} but requires $N_{\mathrm{Lanczos}} \times N_{\mathrm{pol}}$ times applications of the Liouville operator, where $N_{\mathrm{Lanczos}}$ is the length of the Lanczos recursive chain, and $N_{\mathrm{pol}}$ is the number of polarization directions. In figure~\ref{fig:scaling}(b), $N_{\mathrm{Lanczos}}$ is 1600, nearly two orders of magnitude larger than $N_{\mathrm{Davidson}}$. Hence, in the case of Si, the time spent on I/O operations is just a small fraction of the total time to compute the spectrum.

\section{Vertical excitation energies of spin defects}
\label{sec:applications}

The NV$^-$ and neutral silicon-vacancy (SiV$^0$) in diamond, and the neutral divacancy (VV$^0$) in SiC are spin defects in wide-gap materials with numerous applications in quantum technologies~\cite{nv_doherty_2013,nv_schirhagl_2014,nv_gali_2019,nv_barry_2020,siv_rose_2018,siv_green_2019,sic_atature_2018,sic_awschalom_2018,sic_zhang_2020,sic_son_2020}. We report the VEEs of the low-lying excited states of these spin defects, computed using G$_0$W$_0$-BSE@PBE as implemented in the WEST-BSE code. For each system, the atomic positions in the ground state were optimized using the PBE functional until the forces acting on the atoms are smaller than $10^{-3}$ Ry/Bohr and the total energy change between two consecutive iterations is smaller than $10^{-4}$ Ry.

\subsection{Nitrogen-vacancy center in diamond}
\label{sec:nv_diamond}

\begin{figure}[ht!]
\includegraphics[width=0.45\textwidth]{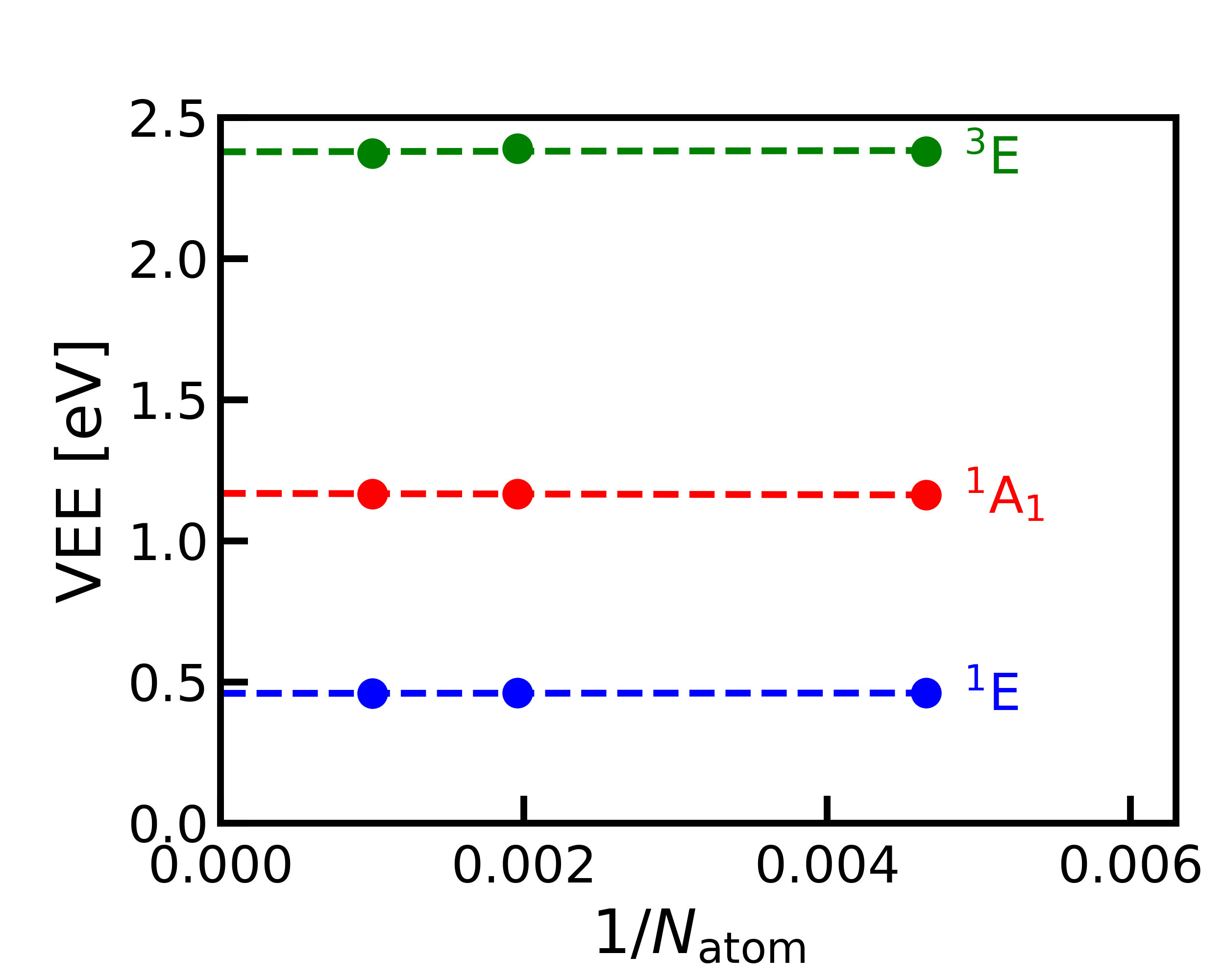}
\caption{Computed vertical excitation energies (VEEs) from the ground state to the $^3E$, $^1A_1$, and $^1E$ excited states of the nitrogen-vacancy center in diamond as a function of the number of atoms ($N_{\mathrm{atom}}$) in the supercell. VEEs are obtained with GW-BSE. Dashed lines show linear extrapolations of the VEEs as a function of $1/N_{\mathrm{atom}}$.}
\label{fig:nv_vee}
\end{figure}

\begin{figure*}[ht!]
\includegraphics[width=0.8\textwidth]{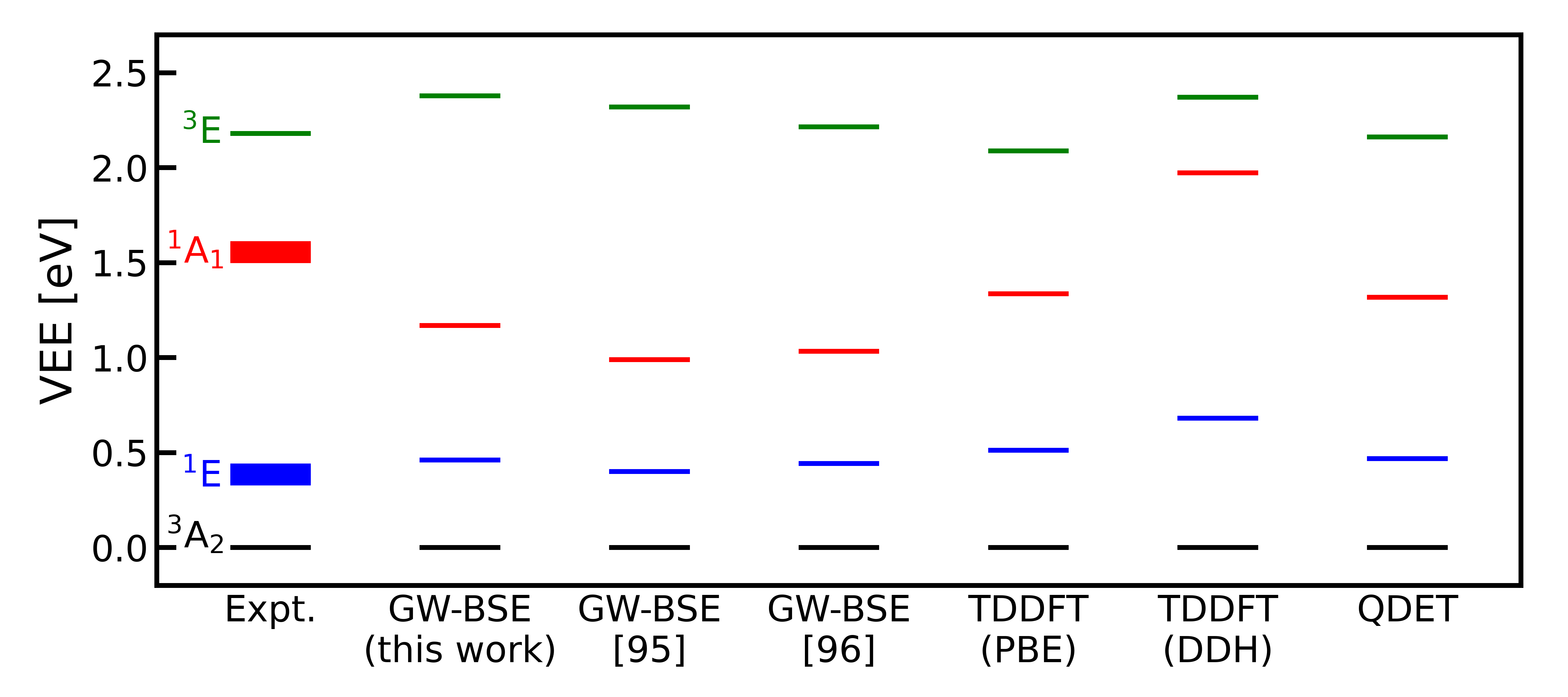}
\caption{Computed vertical excitation energies (VEEs) of the nitrogen-vacancy center in diamond. We also report previous results obtained with GW-BSE~\cite{bse_ma_2010,bse_barker_2022}, the quantum defect embedding theory (QDET)~\cite{qdet_sheng_2022}, and time-dependent density functional theory (TDDFT)~\cite{tddft_jin_2023} with the PBE and DDH functionals, and experimentally inferred VEE and zero-phonon absorption energies of the $^1A_1$ and $^1E$ states~\cite{nv_davies_1976,nv_rogers_2008,nv_goldman_2015a,nv_goldman_2015b}. The thick bars of the $^1A_1$ and $^1E$ states correspond to the range of the experimental results.}
\label{fig:nv_method}
\end{figure*}

The NV$^-$ in diamond has $C_{3v}$ symmetry, with the $a_1$ and the degenerate $e_x$ and $e_y$ single-particle defect states located within the band gap of diamond. In the $m_S = 1$ sublevel of the triplet ground state $^3A_2$, the $a_1^{\uparrow}$, $e_x^{\uparrow}$, $e_y^{\uparrow}$, and $a_1^{\downarrow}$ states are occupied, while the $e_x^{\downarrow}$ and $e_y^{\downarrow}$ states are empty. Spin-conserving excitations from $a_1^{\downarrow}$ to $e_x^{\downarrow}$ and $e_y^{\downarrow}$ yield a triplet excited state $^3E$. Spin-flip excitations from $e_x^{\uparrow}$ and $e_y^{\uparrow}$ to $e_x^{\downarrow}$ and $e_y^{\downarrow}$ yield singlet excited states $^1E$ and $^1A_1$. The transitions between these singlet states play an essential role in the operation of the NV$^-$ centers as qubits, especially in the initialization and readout of the qubit state~\cite{nv_doherty_2011,nv_maze_2011,nv_thiering_2018a,nv_thiering_2018b}. We computed the VEEs of the triplet and singlet excited states of the NV$^-$ in diamond using spin-conserving and spin-flip BSE, respectively. To estimate finite-size effects on the VEEs, we performed calculations with $3 \times 3 \times 3$, $4 \times 4 \times 4$, and $5 \times 5 \times 5$ supercells, corresponding to a number of atoms $N_{\mathrm{atom}}$ equal to 215, 511, and 999, respectively. The results are reported in figure~\ref{fig:nv_vee}. In general, the VEEs of the NV$^-$ in diamond exhibit a weak dependence on the supercell size, since the excitations considered here mainly involve transitions between highly localized defect states within the band gap. The VEEs obtained with the $5 \times 5 \times 5$ supercell are 0.460, 1.167, and 2.373 eV for the $^1E$, $^1A_1$, and $^3E$ excited states, respectively, and are compared with the results of other methods and experiments~\cite{nv_davies_1976,nv_rogers_2008,nv_goldman_2015a,nv_goldman_2015b} in figure~\ref{fig:nv_method}.

Compared to the experimental data, our GW-BSE results show a modest overestimation of the VEE of the $^3E$ triplet state by $\sim$0.2 eV, and an underestimation of the VEE of the $^1A_1$ singlet state by $\sim$0.4 eV. This discrepancy may be attributed to the use of the TDA, the neglect of dynamical screening effects~\cite{bse_rocca_2010}, and contributions of double excitations to the singlet states~\cite{tddft_jin_2022}, which are absent in GW-BSE. Additionally, it has been shown that quantum vibronic coupling affects the position of the VEEs in diamond~\cite{vibronic_kundu_2024}. However, these effects were not included here as a detailed comparison of our results with experiments is beyond the scope of the present study. Our GW-BSE results are in good agreement with previous GW-BSE data~\cite{bse_ma_2010,bse_barker_2022}, differing by less than 0.2 eV. The minor discrepancy may be attributed to different numerical approximations employed here and in previous works. Ma et al.~\cite{bse_ma_2010} employed a 256-atom supercell and they did not mention the code they used. Barker et al.~\cite{bse_barker_2022} mentioned that the code used will be available in BerkeleyGW in the future, but they did not mention the supercell size. In addition, Barker et al. utilized the generalized plasmon pole approximation for the frequency-dependent dielectric screening in their GW calculation, whereas we carried out an integration over the full frequency domain. Further, both our work and that of Barker et al. employed the TDA in the BSE calculations, whereas Ma et al. did not. Our GW-BSE results differ by 0.2 eV from QDET results~\cite{qdet_sheng_2022,tddft_jin_2023}, which can be ascribed, at least in part, to the contribution of double excitations taken into account in QDET. The results of GW-BSE@PBE and TDDFT@DDH~\cite{tddft_jin_2023} for triplet states show good agreement (the fraction of the exact exchange $\alpha = 0.18$ in the DDH functional~\cite{ddh_skone_2014,ddh_skone_2016}). However, for the singlet states, GW-BSE predicted lower VEEs than TDDFT@DDH did.

We emphasize here the advantage of having in the same code (WEST) efficient implementations of the BSE, TDDFT, and QDET methods, thereby enabling an unequivocal comparison between these methods. The BSE and QDET codes share the same implementation of G$_0$W$_0$ (\texttt{wstat} and \texttt{wfreq}, see figure~\ref{fig:workflow}). The BSE and TDDFT codes share the same implementation of the Davidson diagonalization algorithm (\texttt{wbse} in figure~\ref{fig:workflow}). In contrast, numerous comparisons of results obtained with different methods, reported in the literature, may suffer from differences between the numerical approximations and implementations of the various methods, preventing an unequivocal comparison between different levels of theory.

\subsection{Silicon-vacancy in diamond}

\begin{figure}[ht!]
\includegraphics[width=0.45\textwidth]{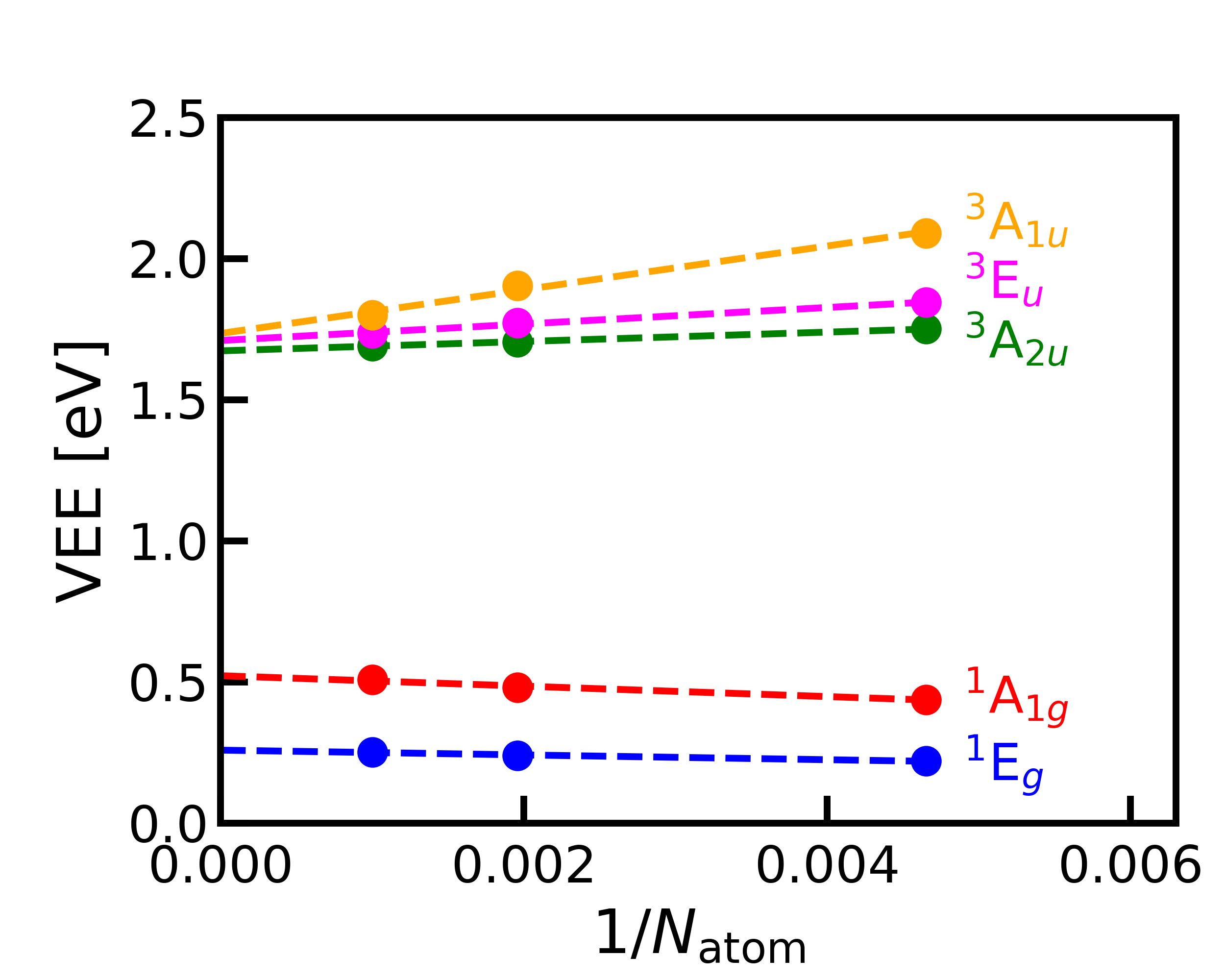}
\caption{Computed vertical excitation energies (VEEs) from the ground to the $^1E_g$, $^1A_{1g}$, $^3A_{2u}$, $^3E_u$, and $^3A_{1u}$ excited states of the silicon-vacancy center in diamond as a function of the number of atoms ($N_{\mathrm{atom}}$) in the supercell. VEEs are obtained with GW-BSE. Dashed lines show linear extrapolations of the VEEs as a function of $1/N_{\mathrm{atom}}$.}
\label{fig:siv_vee}
\end{figure}

The SiV$^0$ in diamond has $D_{3d}$ symmetry, with the $e_g$ single-particle defect states localized in the band gap and the $e_u$ states resonant with the valence bands of diamond. In the $m_S = 1$ sublevel of the triplet ground state $^3A_{2g}$, the $e_{gx}^{\uparrow}$ and $e_{gy}^{\uparrow}$ states are occupied, while the $e_{gx}^{\downarrow}$ and $e_{gy}^{\downarrow}$ states are empty. Spin-conserving excitations from $e_{ux}^{\downarrow}$ and $e_{uy}^{\downarrow}$ to $e_{gx}^{\downarrow}$ and $e_{gy}^{\downarrow}$ yield three triplet excited states $^3A_{2u}$, $^3E_u$, and $^3A_{1u}$, which are all linear combinations of the $e_{ux}^{\downarrow}$ to $e_{gx}^{\downarrow}$, $e_{ux}^{\downarrow}$ to $e_{gy}^{\downarrow}$, $e_{uy}^{\downarrow}$ to $e_{gx}^{\downarrow}$, and $e_{uy}^{\downarrow}$ to $e_{gy}^{\downarrow}$ single excitations. Spin-flip excitations from $e_{gx}^{\uparrow}$ and $e_{gy}^{\uparrow}$ to $e_{gx}^{\downarrow}$ and $e_{gy}^{\downarrow}$ yield singlet excited states $^1E_g$ and $^1A_{1g}$. We computed the VEEs of the triplet and singlet excited states using spin-conserving and spin-flip BSE, respectively. Our calculations were conducted with $3 \times 3 \times 3$, $4 \times 4 \times 4$, and $5 \times 5 \times 5$ supercells of diamond to account for finite-size effects on the VEEs. The results are reported in figure~\ref{fig:siv_vee}. As for the NV$^-$ center in diamond, we find that the VEEs of the singlet excited states depend weakly on the supercell size, since the corresponding excitations mainly involve transitions between highly localized defect states within the band gap. The VEEs obtained with the $5 \times 5 \times 5$ supercell are 0.259 and 0.523 eV for the $^1E_g$ and $^1A_{1g}$ states, respectively. Instead, the VEEs of the triplet excited states, particularly $^3E_u$ and $^3A_{1u}$, exhibit a linear dependence on $1/N_{\mathrm{atom}}$, since the corresponding excitations involve transitions from $e_u$ to $e_g$, with the former being hybridized with the valence bands of diamond. This observation is consistent with the findings of our TDDFT calculations~\cite{tddft_jin_2023}. We extrapolated the VEEs of the $^3A_{2u}$, $^3E_u$, and $^3A_{1u}$ states to the dilute limit, obtaining 1.674, 1.710, and 1.735 eV respectively.

\subsection{Divacancy in 3C silicon carbide}

Similar to the NV$^-$ in diamond, the VV$^0$ in 3C SiC has $C_{3v}$ symmetry with $a_1$ and $e$ defect states. While the $1e^{\uparrow}$ and $a_1^{\downarrow}$ states are located within the band gap of 3C SiC, the $1e^{\downarrow}$ states are resonant with the conduction bands of the host, due to the narrower band gap of 3C SiC compared to that of diamond. The lowest-energy spin-conserving excitation of the VV$^0$ in 3C SiC involves defect-to-defect and defect-to-bulk transitions~\cite{sic_gao_2022}. We computed the VEE of the lowest-energy excited state of the VV$^0$ in 3C SiC. Our calculations were conducted with $3 \times 3 \times 3$, $4 \times 4 \times 4$, and $5 \times 5 \times 5$ supercells of 3C SiC to account for finite-size effects. We obtained 1.414, 1.453, and 1.465 eV, respectively, and observed a linear dependence of the energies on $1/N_{\mathrm{atom}}$. Extrapolating the VEE to the dilute limit yields a value of 1.478 eV. The VEE obtained using the $5 \times 5 \times 5$ supercell is 1.465 eV, which is $\sim$0.4 eV higher than the GW-BSE result reported by Gao et al. using the same supercell~\cite{sic_gao_2022}, and $\sim$0.1 eV higher than the value from our TDDFT calculations using the DDH ($\alpha = 0.15$) functional. By optimizing the excited-state geometry using TDDFT analytical forces~\cite{tddft_jin_2023} and the PBE functional, we obtained the Franck-Condon shift of the excited state, which equals 0.15 eV. Subtracting this energy from the GW-BSE VEE yields 1.32 eV, which is 0.2 eV higher than the measured zero-phonon line (ZPL) of 1.12 eV~\cite{sic_falk_2013} (note that, as for diamond, we did not include here quantum vibronic effects). For the lowest ten excited states, we computed the radiative lifetime of the $s$-th excited state $t_s$ as~\cite{lifetime_davidsson_2020}
\begin{equation}
t_s = \frac{3 \pi \epsilon_0 c^3 \hbar^4}{n | \omega_s |^3 | \mathbf{\mu}_s |^2} \,,
\end{equation}
where $\mathbf{\mu}_s$ is the transition dipole moment between the ground and the excited state, $\epsilon_0$ is the vacuum permeability, $c$ is the speed of light in vacuum, and $n = 2.65$ is the refractive index of 3C SiC. The lowest-energy excited state has the shortest radiative lifetime, 25 ns, a value close to the experimental estimate of 23 ns~\cite{sic_christle_2017}.

\begin{figure}[ht!]
\includegraphics[width=0.9\textwidth]{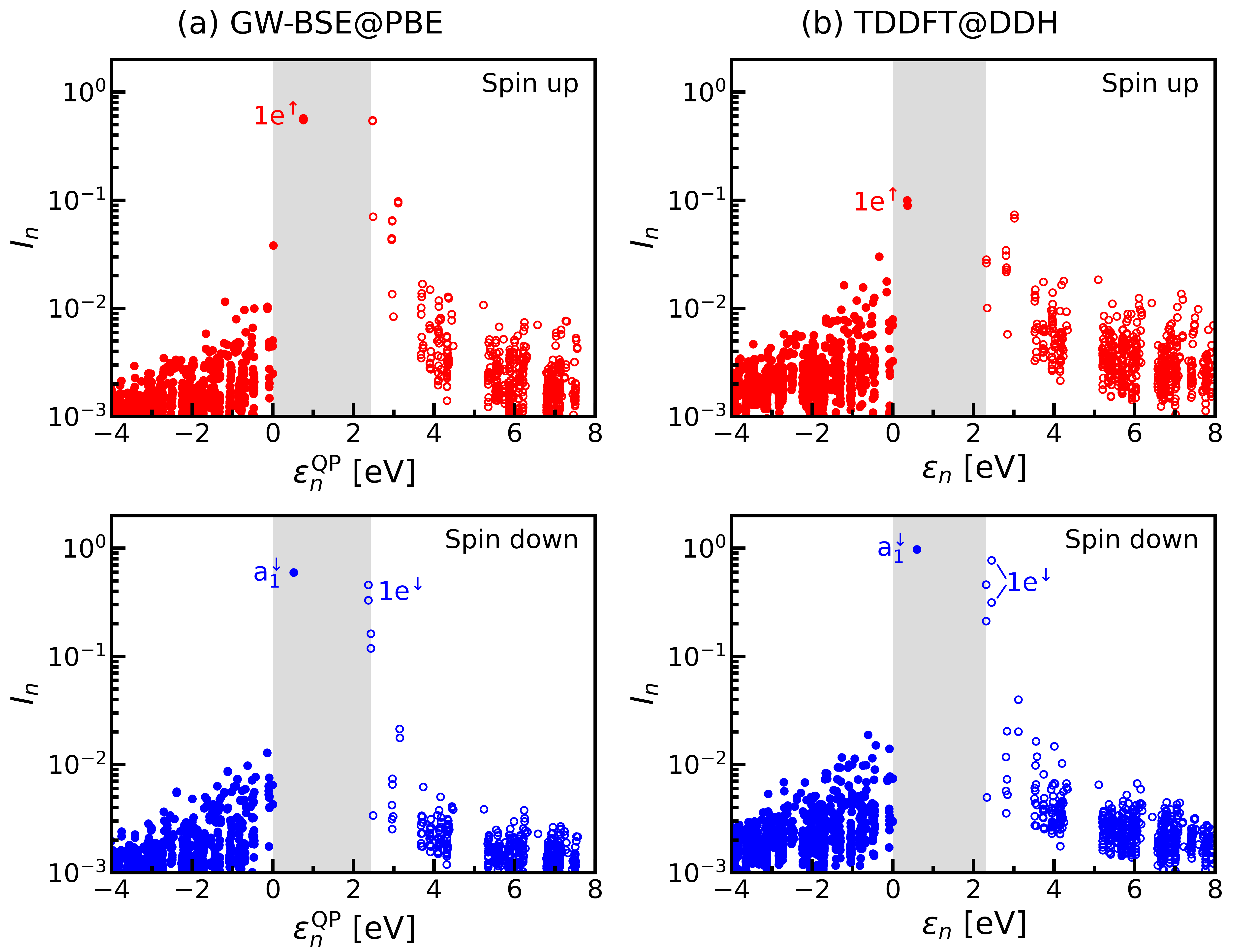}
\caption{Contribution of Kohn-Sham (KS) wave functions in spin up (red) and spin down (blue) channels to the lowest-energy, multireference excited state of the divacancy center in 3C silicon carbide ($5 \times 5 \times 5$ supercell). The results were obtained with (a) GW-BSE using the PBE functional, plotted as a function of the quasi-particle energy $\varepsilon_n^{\mathrm{QP}}$, and (b) time-dependent density functional theory (TDDFT) using the DDH functional ($\alpha = 0.15$), plotted as a function of the KS energies $\varepsilon_n$. The function $I_n$ is defined in equation~\ref{eq:factor}. Filled and open symbols denote occupied and empty states, respectively. The shaded area indicates the band gap of the material.}
\label{fig:vv_exc}
\end{figure}

To understand the character of the lowest-energy excited state, we project its wave function ($A_{s = 1} = \{ \ket{a_{1,v}}: v = 1, \dots, N_{\mathrm{occ}} \}$) onto the empty KS wave functions $\{ \ket{\psi_c}: c = N_{\mathrm{occ}} + 1, \dots, N_{\mathrm{occ}} + N_{\mathrm{empty}}\}$,
\begin{equation}
c_{vc} = \braket{a_{1,v} | \psi_c} \,.
\end{equation}
We then define the quantity
\begin{equation}
\label{eq:factor}
I_n = \left( \sum_{m} | c_{nm} |^2 \right)^{1/2} \,,
\end{equation}
as a measure of the contribution of the $n$-th band to $A_{s = 1}$. The summation over $m$ runs from $N_{\mathrm{occ}} + 1$ to $N_{\mathrm{occ}} + N_{\mathrm{empty}}$ if the $n$-th band is occupied, and from 1 to $N_{\mathrm{occ}}$ if the $n$-th band is empty. In figure~\ref{fig:vv_exc}(a), we plot $I_n$ as a function of the QP energy $\varepsilon_v^{\mathrm{QP}}$. The lowest excitation found here primarily involves transitions in the spin up channel from the $1e^{\uparrow}$ defect state and the valence band maximum (VBM) of 3C SiC to the conduction band minimum (CBM), and transitions in the spin down channel from the $a_1^{\downarrow}$ defect state to the $1e^{\downarrow}$ defect state and the CBM. This result is consistent with the analysis reported by Gao et al.~\cite{sic_gao_2022}. In figure~\ref{fig:vv_exc}(b), we plot $I_n$ obtained using TDDFT and the DDH functional and we compare it with that computed with GW-BSE. Both methods predict a similar composition of the excited states involved in the transition, with the main difference being a smaller contribution from the spin up channel in the TDDFT calculations.

\begin{figure}[ht!]
\includegraphics[width=0.45\textwidth]{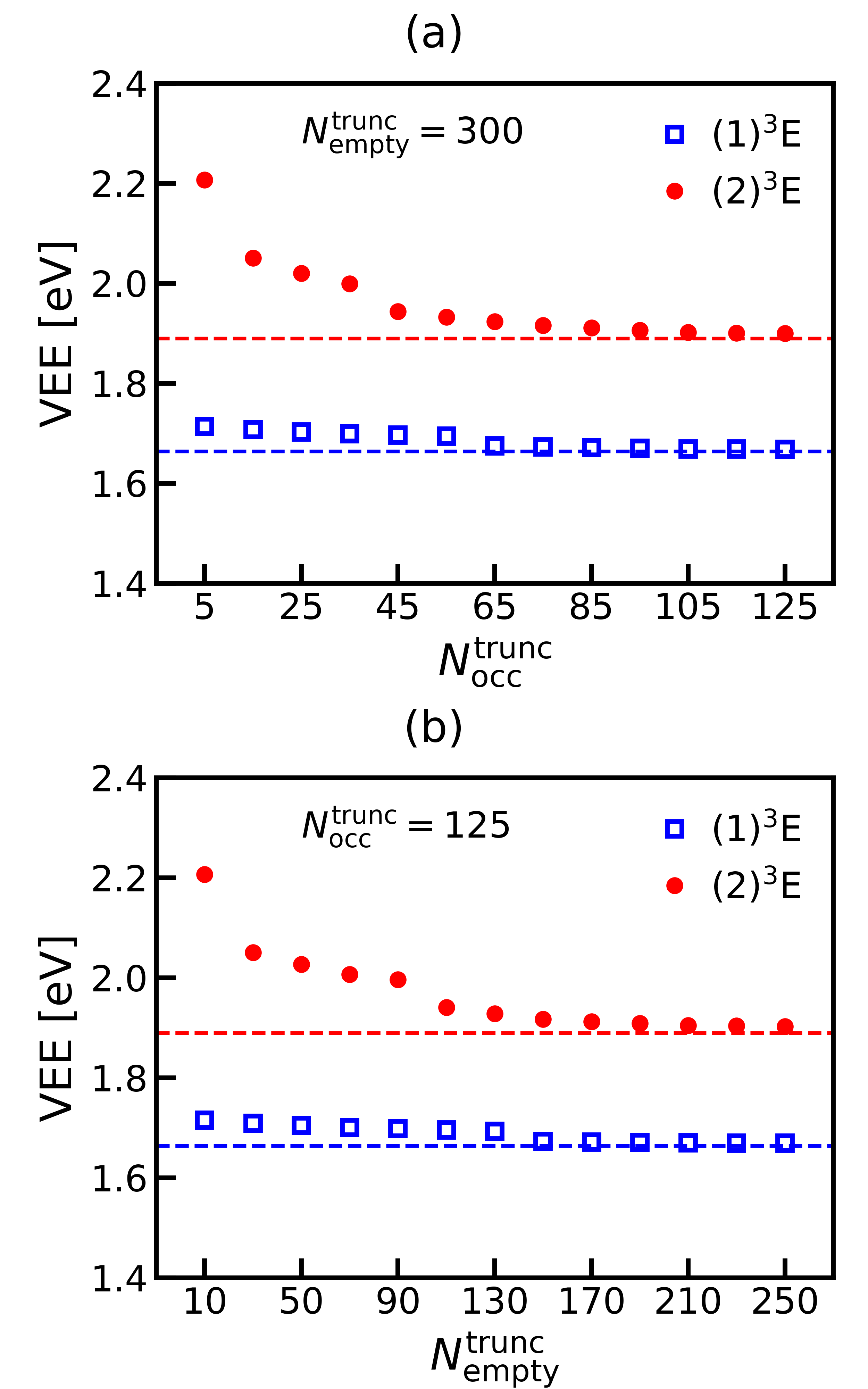}
\caption{Vertical excitation energies (VEEs) from the ground to the excited states of the divacancy center in 3C silicon carbide, computed with time-dependent density functional theory (TDDFT) using an electron-hole (e-h) basis set, the DDH functional ($\alpha = 0.15$), and a $2 \times 2 \times 2$ supercell. The blue squares and red circles represent the VEEs of the (1)$^3E$ and (2)$^3E$ excited states, respectively. (a) Convergence with respect to the number of occupied states ($N_{\mathrm{occ}}^{\mathrm{trunc}}$) in the e-h basis set, with the number of empty states ($N_{\mathrm{empty}}^{\mathrm{trunc}}$) fixed at 300. (b) Convergence with respect to $N_{\mathrm{empty}}^{\mathrm{trunc}}$, with $N_{\mathrm{empty}}^{\mathrm{trunc}}$ fixed at 125, including all the occupied states. The dashed lines indicate the values obtained using the formulation presented in section~\ref{sec:theory}, without explicitly computing empty states.}
\label{fig:vv_conv}
\end{figure}

We note that the GW-BSE calculation by Gao et al.~\cite{sic_gao_2022} restricted the solution of the BSE to an energy window of $[$VBM $-$ 0.5 eV, CBM $+$ 1.6 eV$]$ (40 occupied bands and 40 empty bands for a system of 998 atoms). As shown in figure~\ref{fig:vv_exc}, bands outside this range collectively yield a small but non-negligible contribution to the lowest-energy excited state. In figure~\ref{fig:vv_conv}, we present the VEEs of two representative excited states of the VV$^0$ in 3C SiC, labeled as (1)$^3E$ and (2)$^3E$ since both states have the $^3E$ symmetry. Using TDDFT with an e-h basis set, the DDH functional ($\alpha = 0.15$), and a $2 \times 2 \times 2$ supercell, we tested the convergence of the VEEs as a function of the number of occupied ($N_{\mathrm{occ}}^{\mathrm{trunc}}$) and empty ($N_{\mathrm{empty}}^{\mathrm{trunc}}$) bands in the e-h basis set. Our results indicate that a large energy window of $[$VBM $-$ 3 eV, CBM $+$ 6 eV$]$ (35 occupied bands and 90 empty bands for a system with 62 atoms) is necessary to converge the VEE of the (2)$^3E$ excited state to a precision of 0.1 eV. The VEE of the (1)$^3E$ excited state converges faster with respect to $N_{\mathrm{occ}}^{\mathrm{trunc}}$ and $N_{\mathrm{empty}}^{\mathrm{trunc}}$. The VEEs of both states eventually converge to the values obtained using the formulation presented in section~\ref{sec:theory} and the WEST code. We emphasize that the formulation based on the linearized Liouville equation and density matrix perturbation theory has enabled us to include all occupied bands explicitly and all empty bands implicitly, thus providing an accurate description of the excited states, which is challenging to achieve in conventional BSE or TDDFT implementations relying on the use of an e-h basis set.

\subsection{Nitrogen-vacancy center at a dislocation core in diamond}

We now turn to describing the electronic structure of spin defects in the presence of dislocations. The latter are prevalent extended defects in diamond, leading to the breaking of the long-range translational symmetry along the dislocation line. Due to the lower formation energies of NV centers at the dislocation core compared to the bulk, it has been suggested that a chain of NV centers could form along the dislocation line~\cite{dislocation_ghassemizadeh_2022}. Hence, for numerous applications in quantum technologies, it is important to determine the electronic structure of NV centers located at a dislocation core and compare it to that of NV centers in bulk diamond. The GPU-accelerated solver of GW-BSE is used here to compute the VEEs of a system composed of a point and an extended defect, which requires complex supercells with over a thousand atoms.

\begin{figure}[ht!]
\includegraphics[width=0.9\textwidth]{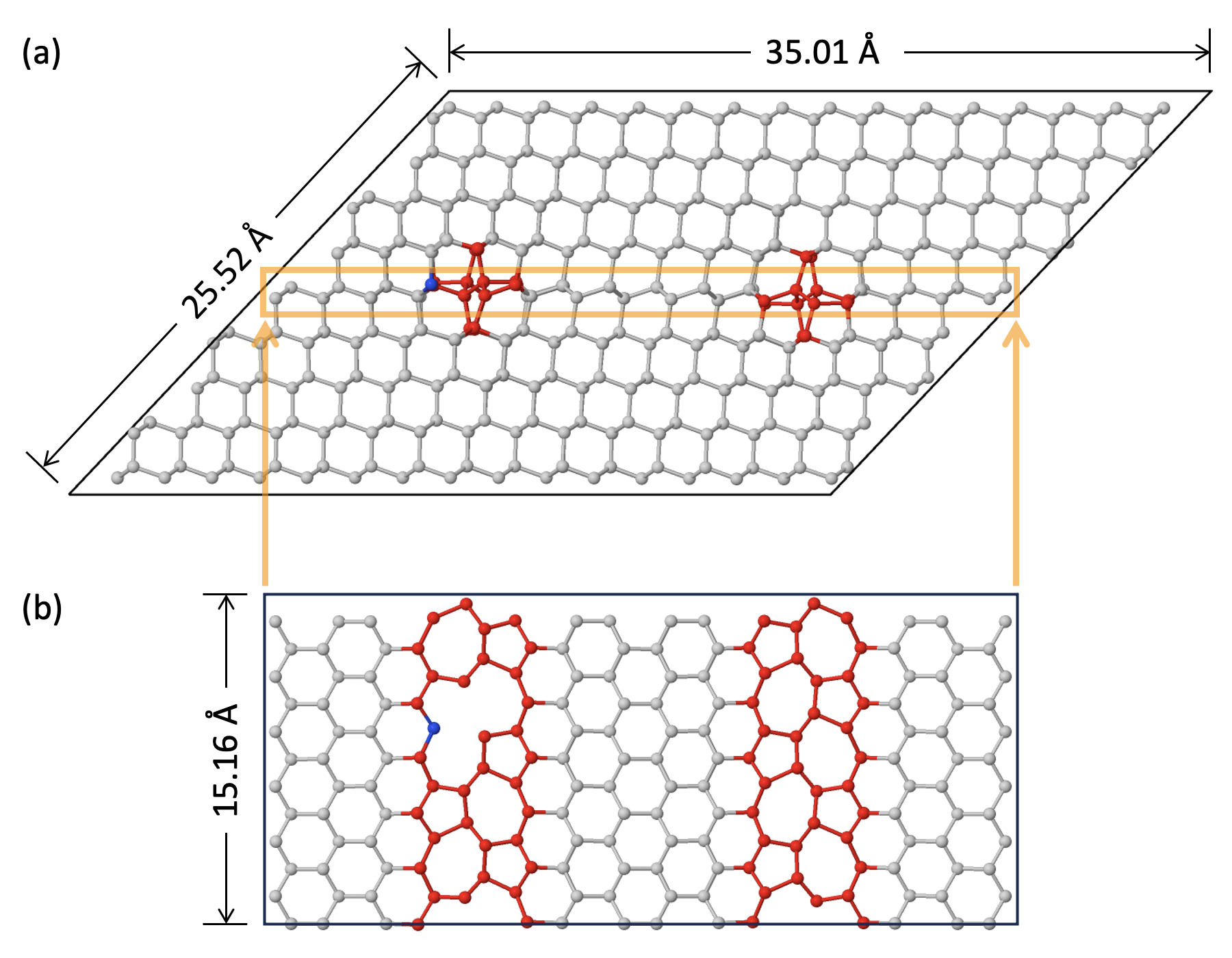}
\caption{Supercell used to represent a nitrogen-vacancy center at a 90$^{\circ}$ double period partial glide dislocation core in diamond. The supercell includes two dislocation cores oriented in opposite directions. Carbon atoms at the dislocation cores are shown in red, while the rest of the carbon atoms are shown in gray. The nitrogen atom is shown in blue. (a) Top view along the dislocation lines, with the glide plane highlighted by the orange box. (b) Side view of the glide plane, corresponding to the atoms within the orange box in panel (a). In panel (b), atoms outside the orange box in panel (a) are omitted for clarity.}
\label{fig:nv_disloc}
\end{figure}

We focused on a specific NV configuration located at the core of a 90$^{\circ}$ double-period partial glide dislocation, as illustrated in figure~\ref{fig:nv_disloc}. Similar to the case of NV$^-$ in bulk diamond, this NV configuration exhibits a triplet ground state and is stable in its $-1$ charge state. The supercell model includes two dislocation cores (highlighted by red atoms in figure~\ref{fig:nv_disloc}) and comprises a total of 1727 atoms and 6910 electrons. As discussed in section~\ref{sec:nv_diamond}, the NV$^-$ in bulk diamond features two degenerate triplet excited states $^3E$, arising from spin-conserving excitations from the $a_1^{\downarrow}$ to the $e_x^{\downarrow}$ and $e_y^{\downarrow}$ states. The GW-BSE VEE for the $^3E$ states is 2.38 eV. For the NV$^-$ located at the dislocation core, the degeneracy of the $e_x^{\downarrow}$ and $e_y^{\downarrow}$ states is lifted, resulting in a splitting of over 0.5 eV. Hence, spin-conserving excitations from $a_1^{\downarrow}$ to $e_x^{\downarrow}$ and $e_y^{\downarrow}$ yield two distinct triplet excited states, with GW-BSE VEEs of 2.08 and 2.76 eV, respectively. Our study of the NV$^-$ center at a dislocation core in diamond represents the first demonstration of the capability to explore the interaction between a point defect and an extended defect within a large supercell, at the MBPT level of theory.

\section{Conclusions}
\label{sec:conclusions}

In this article, we presented the use of GPUs to carry out large-scale GW-BSE calculations of heterogeneous systems with the WEST code. Compared to conventional implementations of the BSE, the algorithms implemented in WEST eliminate the need for computing explicitly empty states and inverting large dielectric matrices. We outlined a series of optimizations that enhance the performance of the code on GPU-equipped supercomputers, including a multilevel parallelization scheme, strategies to mitigate the overhead cost associated with MPI communications, and a GPU-accelerated Wannier localization technique based on the JADE algorithm.

The GPU version of WEST-BSE is the first massively parallel implementation of the BSE on GPUs, achieving excellent scalability, as demonstrated on the Perlmutter supercomputer, where we used up to 4096 NVIDIA A100 GPUs. We showcased the performance of the code through calculations of the absorption spectrum of bulk silicon and vertical excitation energies of spin defects in wide-gap materials, carried out with supercells with up to 6910 electrons. Importantly, the WEST-BSE code allows for detailed finite size scaling studies and for the extrapolation of the results for VEEs to the dilute limit. For the first time, we used GW-BSE calculations to study the nitrogen-vacancy center at a dislocation core in diamond, illustrating the capability to investigate the interaction between a point defect and an extended defect.

We are exploring the possibility of extending the WEST-BSE code to solve the full BSE without the TDA~\cite{bse_rocca_2010}, and to include spin-orbit coupling effects, targeting materials that contain heavy elements. Additionally, we are experimenting with the use of single precision in WEST-BSE to achieve higher performance on GPUs~\cite{west_yu_2022}.

Data related to this publication are organized using the Qresp software~\cite{qresp_govoni_2019} and are available online at \url{https://paperstack.uchicago.edu}.

The authors declare no competing financial interest.

\begin{acknowledgement}
The authors acknowledge support provided by the Midwest Integrated Center for Computational Materials (MICCoM), as part of the Computational Materials Sciences Program funded by the U.S. Department of Energy, Office of Science, Basic Energy Sciences, Materials Sciences, and Engineering Division through Argonne National Laboratory. This research used resources of the National Energy Research Scientific Computing Center (NERSC), a U.S. Department of Energy Office of Science User Facility operated under Contract No. DE-AC02-05CH11231. This research also used resources of the Argonne Leadership Computing Facility at Argonne National Laboratory, a U.S. Department of Energy Office of Science User Facility operated under Contract No. DE-AC02-06CH11357. The authors thank Maryam Ghazisaeidi, Sevim Polat Genlik, and Cunzhi Zhang for providing the atomic structure of the nitrogen-vacancy center at a dislocation core in diamond.
\end{acknowledgement}

\bibliography{bse}

\end{document}